\documentclass[]{aa}
\usepackage{graphicx}
\usepackage{times}
\usepackage{natbib}
\bibpunct{(}{)}{;}{a}{}{,} 
\let\bibfont=\small

\newcommand{\kms}{km~s$^{-1}$}
\newcommand{\myr}{$\dot{M}$~yr$^{-1}$}
\begin{document}
  \title{Stellar wind bubbles around WR and [WR] stars}

  \author{Garrelt Mellema\inst{1} \and Peter Lundqvist\inst{2}}

  \institute{Sterrewacht Leiden, P.O. Box 9513, 2300 RA, Leiden, 
             The Netherlands\\
             \email{mellema@strw.leidenuniv.nl}
           \and
             Stockholm Observatory, AlbaNova (SCFAB), Department of Astronomy,
                     SE-106 91 Stockholm, Sweden\\
            \email{peter@astro.su.se}}
  \offprints{G.~Mellema}

  \date{Received 30 April 2002 / Accepted 9 August 2002}

  \abstract{We study the dynamics of stellar wind bubbles around
  hydrogen-deficient stars using numerical simulations with time- and ion
  dependent cooling. We consider two types of hydrogen-deficient stars,
  massive WR stars, producing Ring Nebulae, and low mass [WR] stars, producing
  Planetary Nebulae. We show that for the Planetary Nebulae, the different
  cooling properties of the hydrogen-deficient wind lead to a later
  transition from momentum- to energy-driven flow, which could explain the
  observed turbulence of these nebulae. We find that Ring Nebulae should all
  be energy-driven, and show how comparing the bubble's momentum and kinetic 
  energy to the input wind momentum and kinetic energy, can give misleading
  information about the dynamics of the bubble.

 }

  \maketitle
%

\section{Introduction}

Both high and low mass stars can under certain circumstances reduce the
hydrogen content of their atmospheres.  In most cases this leads to the
so-called Wolf-Rayet (WR) phenomenon, i.e.\ a dense fast wind starting below
the photosphere, which produces a Wolf-Rayet spectrum, dominated by bright
emission lines \citep{AbbottConti87}.

Traditionally most attention was given to the massive WR stars, but the last
ten years their lower mass cousins, the [WR] stars have been studied in more
detail \citep[see e.g.][for a review]{Koesterke01}.

The winds from [WR] and WR stars produce stellar wind bubbles (SWBs) in their
environment. In the case of WR stars they are called Ring Nebulae (RNe), in
the case of [WR] stars, Planetary Nebulae (PNe). PNe also form around H-rich
central stars, so [WR] stars constitute a subgroup among central stars of
PNe. Approximately 7\% of central stars is estimated to be [WR], the rest
being H-rich \citep{Gorny2001} (with the exception of so-called weak emission
line stars or wels which appear to be H-poor without showing the WR
phenomenon). All central stars are considered to be in the same evolutionary
phase, namely the post-AGB phase, where the [WR] have changed their
atmospheric abundances through a timely thermal pulse \citep{Herwig2001}. The
existence of two different groups of central stars suggests that a comparison
between the two could be interesting.

In contrast, WR stars are thought to be a phase in the evolution of most stars
with Zero Age Main Sequence masses higher than $\sim 25$~M$_\odot$, and there
is no class of H-rich stars in an equivalent evolutionary stage. Massive stars
lose mass already on the Main Sequence, followed by a slower wind when the
star moves over to the red part of the Hertzsprung--Russell diagram, and
finally leading to the WR wind. This leads to a whole series of interactions
between the wind phases \citep[see e.g.][] {GuileNorbert1}. The more
complicated environments and probably also the clumpiness of the actual winds
make that the RNe are mostly irregular and filamentary, lacking the overall
symmetries found in PNe.

In this paper we investigate whether the fact that the winds from WR and [WR]
stars are H-poor changes the dynamics of their SWBs. In the case of the [WR]
stars this is relevant because we can compare the PNe between the H-rich and
H-poor central stars. In the case of the WR stars this is relevant because it
has been suggested that even at high wind velocities their SWBs can be
strongly cooling.

The layout of the paper is as follows. The effects of cooling on SWBs are
outlined in Sect.~2.  We investigate the effects of WR winds using numerical
hydrodynamic models with detailed cooling, described in Sect.~3. Section 4
and 5 contain the results of the simulations for PNe and RNe,
respectively. We discuss these results further in Sect.~6 and sum up the
conclusions in Sect.~7.

\section{Stellar wind bubbles}

Stellar wind bubbles (SWBs) expanding into a dense environment (as is the
case in both RNe and PNe), come in two basics types: radiative (also called
momentum-driven or momentum-conserving) and non-radiative (also called
adiabatic or energy-driven/energy-conserving). Whether the SWB is radiative
or non-radiative depends on the cooling at the wind shock (or inner
shock). This is a well-known phenomenon, see for example
\citet{DysonWilliams} or \citet{KooMcKee1,KooMcKee2}.

Since cooling strongly depends on the metallicity, it is expected that the
radiative cooling rate of the metal-rich WR winds will be orders of magnitude
above normal (`solar abundance') cooling rates. This could then lead to SWBs
remaining radiative when they would normally be non-radiative.

Whether a SWB is radiative or non-radiative depends on three time scales
\citep{KooMcKee2}: the crossing time for the free wind $t_{\rm
cross}=R_{\rm sw}/v_{\rm w}$, the age of the bubble $t$, and the cooling time
$t_{\rm cool}$. The cooling time can be found from
\begin{equation}
  t_{\rm cool}={C_1 v_{\rm sw}^3 \over \rho_{\rm pre}}
\end{equation}
where $C_1$ is a constant which describes the cooling, which is normally
taken to be $6.0\times 10^{-35}$~g~cm$^{-6}$~s$^4$. If $t_{\rm cool} \ll
t_{\rm cross}$ the bubble is radiative, if $t \ll t_{\rm cool}$ the bubble is
non-radiative, and if $ t_{\rm cross} \ll t_{\rm cool} \ll t$ cooling does 
affect the shocked wind, but the bubble is still mostly filled with hot 
shocked gas, and the bubble is partially radiative.

The application of these expressions lead to evolutionary sequences for
stellar wind bubbles, as explored by \citet{KooMcKee2} for the most general
case, by \citet{GarciaSeguraMacLow95a} for RNe, and by
\citet{KahnBreitschwerdt} for PNe. Here we limit ourselves to the interaction
of constant or evolving winds interacting with a previous slow wind phase,
the generic model for both PNe and RNe. When a volume of hot shocked
wind material is present, we will call the bubble energy-driven, if not, we
will call it momentum-driven. This means that a partially radiative bubble
will also be considered to be energy-driven.

\section{Numerical method}
We study the formation of SWBs using a numerical hydrodynamics code coupled
to a detailed atomic physics module, which calculates the
ionization/recombination and cooling for the ions included. The hydrodynamic
Euler equations are solved in one spatial dimension using the Roe-solver
\citep{Roe81,Mellemaetal91}. The spatial coordinate is radius. The Roe
solver solves the equations conservatively, and with second order accuracy in
both space and time. The ionic/atomic concentrations are passively advected
using the approach from CLAWPACK \citep{Clawpack}. This advection scheme was
tested using the test problems described in \citet{PlewaMuller} and found to
perform as well as the approach described by those authors.

\subsection{Atomic physics module}
Radiative cooling is due to excitation and ionization of molecules,
atoms and ions, and therefore depends on the composition of the
gas. Most numerical simulations which include cooling, use a so-called
cooling curve. This is a curve which gives cooling as a function of
temperature. The first cooling curves were constructed in the sixties,
and the one which became standard were done around 1970 by Cox and
collaborators \citep{CoxTucker, CoxDaltabuit71, RaymondCoxSmith}.
These curves are calculated assuming collisional ionization
equilibrium to fix the ion concentrations. Although the original
publications contained cooling curves for individual elements (but not
ions), the most common use of them has been in the form of one total
cooling curve for standard abundances, such as the one from
\citet{DalgarnoMcCray}. For cosmological applications, H and He only
cooling curves have been constructed; \citet{DopitaSutherland} present
cooling curves for a range of metallicities. All of the above, as well
as our method below, assume an optically thin plasma, in which all of
the radiation escapes.

Primarily motivated by astrophysical situations in which photo-ionization (and
not collisional ionization) determines the ion concentrations, we have
developed numerical approaches in which cooling is calculated for each ion
separately and time-dependently, using the local temperature and density. The
first version of this was described by \citet{RH-PN1} and \citet{Coolingfits},
and a newer version by \citet{Ragaetal} and \citet{Photoclump}.

Here we use an improved version of the same approach as in \citet{Ragaetal},
and \citet{Photoclump}. The improvements are that we now allow for
position-dependent elemental abundances, which is crucial for the problem
studied here.  We also now include all ions of the following elements: H, He,
C, N, O and Ne. Collisional excitation for the highest ionization stages were
taken from \citet{GaetzSalpeter}. We have also included free-free and
free-bound cooling to allow an extension to $10^9$~K using the method of
\citet{GronenschildMewe}. Free-free Gaunt factors are from
\citet{KarzasLatter}, and relativistic free-free corrections from
\citet{Gould80}.

Furthermore, we dropped the special treatment of H and He, which are now
treated in the same way as the other elements. Since abundances can vary, we
let all elements contribute to the electron density. Cooling due to
collisional ionization is currently not included. This will underestimate the
cooling at lower shock velocities. The module as used for the results in this
paper does not contain photo-ionization, since we are primarily interested in
the effects of the H-deficient winds.

Our module does not include any heavier elements than Ne. At the high
temperatures where the elements included become completely ionized, heavier
elements can start to contribute to the cooling. We therefore underestimate
the radiative cooling in the temperature regime $\log_{10}T=6$ to $6.8$,
beyond which the free-free starts to dominate \citep[see
e.g.][]{GaetzSalpeter}. Whether or not this is a serious underestimate of the
cooling depends on the relative abundances of the cooling elements. In our
simulations it will mostly affect the solar abundances case. We will comment
further on this below.

\begin{figure}
\centerline{\includegraphics[width=8cm]{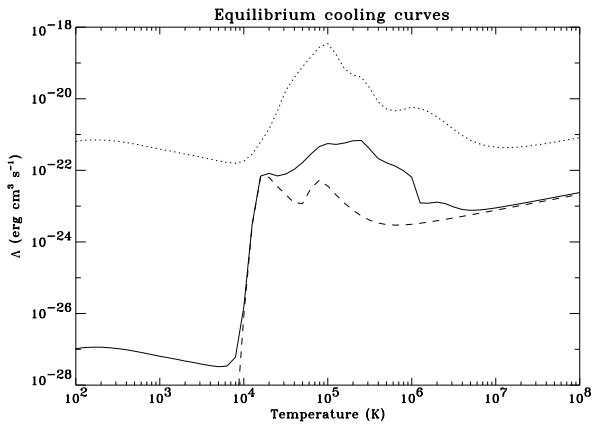}}
\caption{Collisional equilibrium cooling curves calculated with the {\it
DORIC}\/ module. Solid line: solar abundances, dashed line: metal free,
dotted line: [WC] abundances. The curves are {\it not}\/ used in any of the
calculations, but are shown to illustrate the effect of metallicity and the
capability of {\it DORIC}.}
\label{coolingcurves}
\end{figure}

In Fig.~\ref{coolingcurves} we show three equilibrium cooling curves
calculated with the DORIC atomic physics module. One for solar abundances,
one for primordial abundances, and one for [WC] abundances. These curves
illustrate the dramatic effect of abundances on the cooling. Note that
the high cooling rate below $10^4$~K is entirely due to CII, since this
ion is assumed never to recombine (due to a CI ionizing background). 

In the calculations below, we do not use these curves, but instead calculate
the concentrations of ions (and therefore the radiative cooling)
time-dependently. Especially in low density areas the ionic concentrations can
deviate from the equilibrium values. We comment on this in Sect.~5.

\begin{table*}
{Table 1 -- Abundances used in the various simulations}\\
\begin{tabular*}{14cm}{*{7}{l}}
\noalign{\smallskip}
\hline\hline
\noalign{\smallskip}
Element      &  Solar & [WC] & WNl & WNe & WCl & WCe \\
\noalign{\smallskip}
\hline
\noalign{\smallskip}
H            &  $9.21\times 10^{-1}$ & $9.98\times 10^{-3}$ & $1.51\times 10^{-1}$ & $1.01\times 10^{-6}$ & $9.89\times 10^{-5}$ & $1.01 \times 10^{-4}$ \\
He           &  $7.83\times 10^{-2}$ & $7.09\times 10^{-1}$ & $8.38\times 10^{-1}$ & $9.89\times 10^{-1}$ & $6.43\times 10^{-1}$ & $3.83\times 10^{-1}$ \\
C            &  $3.05\times 10^{-4}$ & $2.60\times 10^{-1}$ & $1.01\times 10^{-4}$ & $1.01\times 10^{-4}$ & $2.77\times 10^{-1}$ & $4.64\times 10^{-1}$ \\
N            &  $7.66\times 10^{-5}$ & $9.98\times 10^{-4}$ & $1.01\times 10^{-2}$ & $1.01\times 10^{-2}$ & $9.89\times 10^{-5}$ & $1.01\times 10^{-4}$ \\
O            &  $6.22\times 10^{-4}$ & $2.00\times 10^{-2}$ & $3.03\times 10^{-4}$ & $3.03\times 10^{-4}$ & $7.91\times 10^{-2}$ & $1.51\times 10^{-1}$ \\
Ne           &  $1.11\times 10^{-4}$ & $9.98\times 10^{-4}$ & $1.01\times 10^{-4}$ & $1.01\times 10^{-3}$ & $9.89\times 10^{-4}$ & $1.01\times 10^{-3}$ \\
\noalign{\smallskip}
\hline
\noalign{Crowther, private communication}
\end{tabular*}
\end{table*}

\section{Models for PNe}
The properties of [WR] stars have been well studied, and the results from two
independent groups agree well. For an overview see the reviews of
\citet{Koesterke01} and \citet{Hamann02}.  All [WR] stars are carbon-rich, so
from now on we refer to them as [WC] stars.  What appears to be an
evolutionary sequence of [WC]late, to [WC]early, to the PG~1159 stars, is
well sampled, and the abundances are more or less the same all along this
sequence; a previous discrepancy has now been resolved, (De Marco, private
communication).

The stellar atmosphere analysis yields mass loss rates and velocities. The
results show that the wind velocities are similar to the ones from normal
central stars, but the mass loss rates are $\sim$10 times higher. Therefore,
we will use a higher mass loss rate in our [WC] model.

\begin{figure*}
\centerline{\includegraphics[width=8cm]{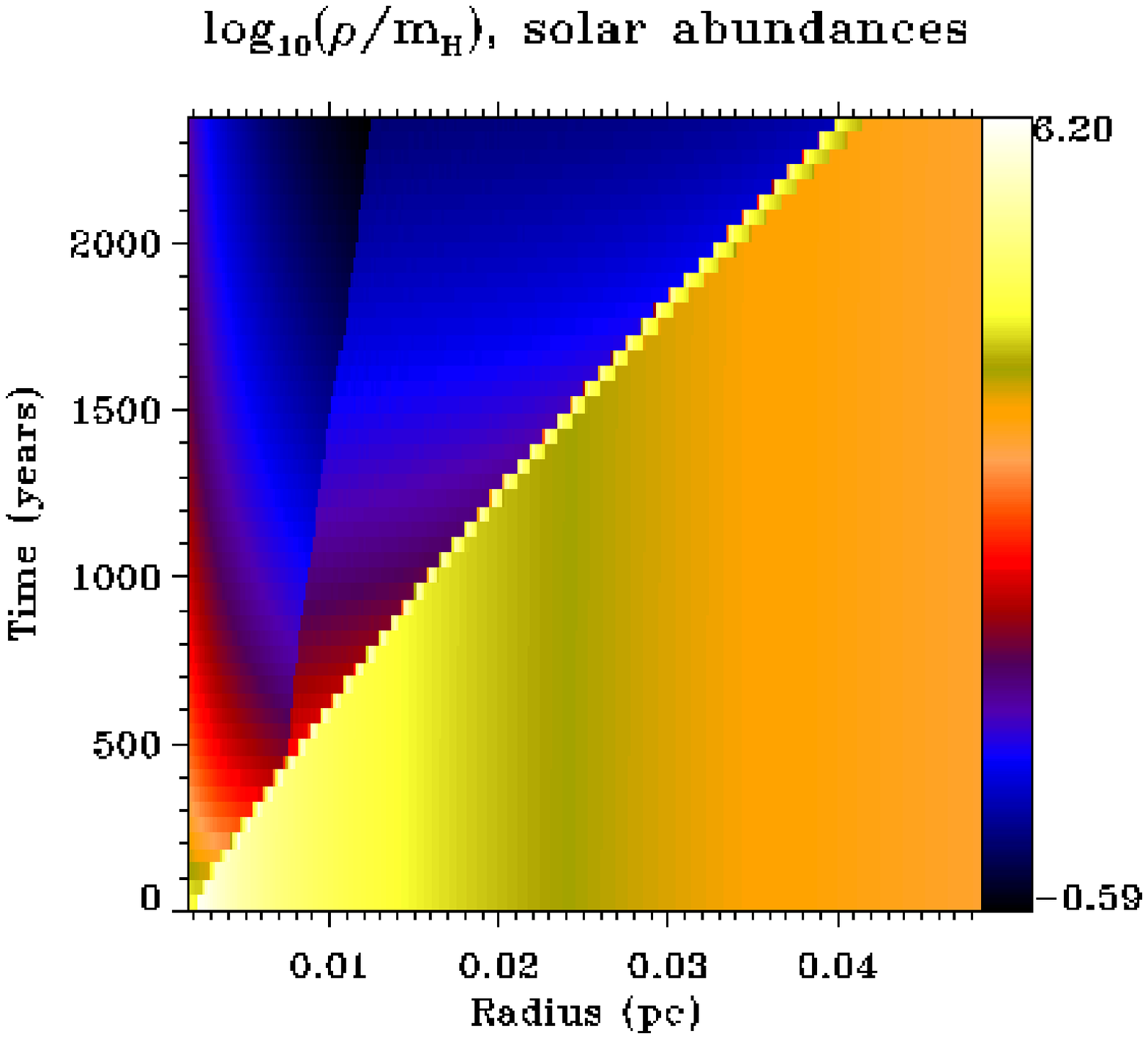}\includegraphics[width=8cm]{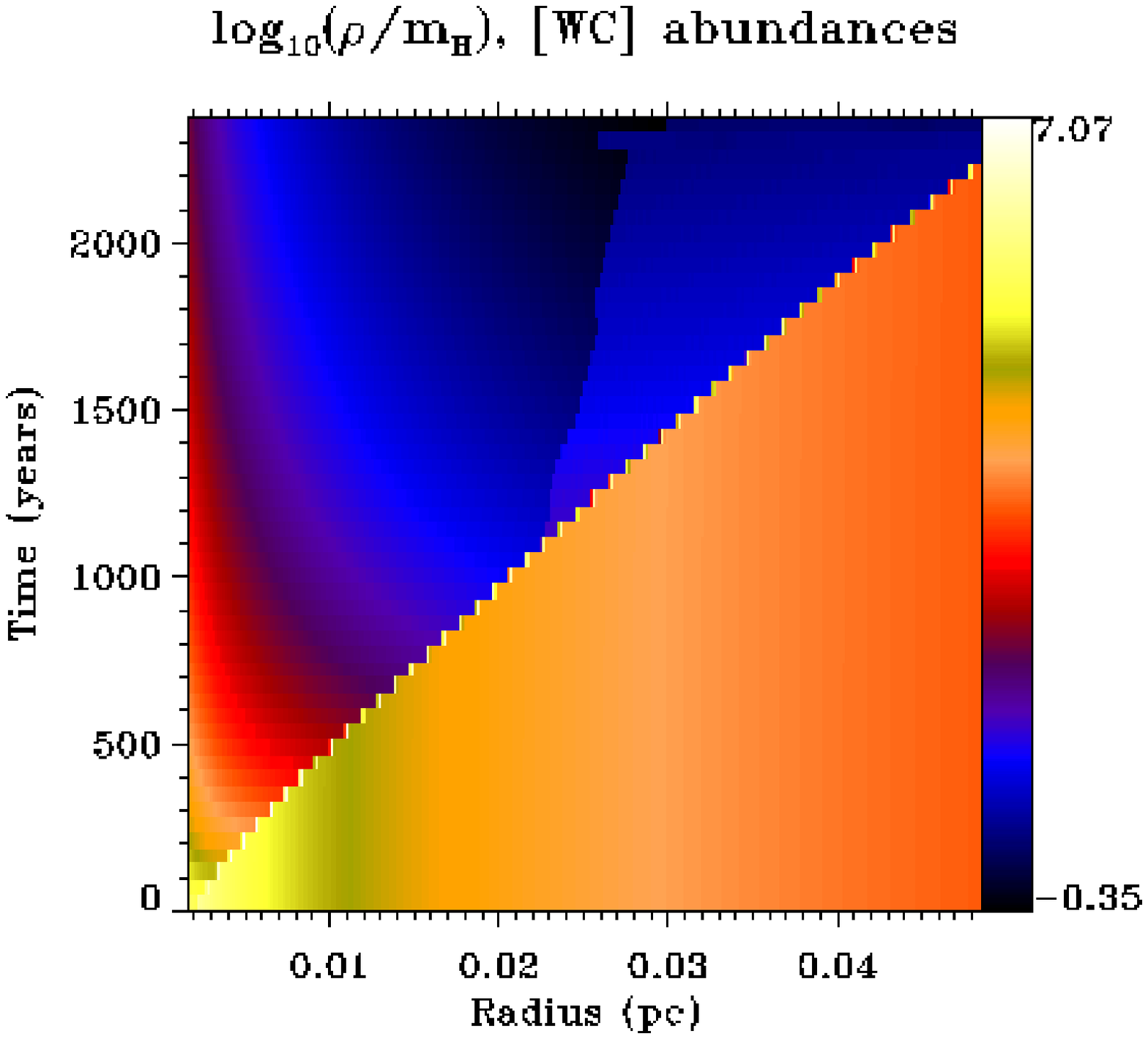}}
\caption{Colour plot of $\log_{10}(\rho /m_{\rm H}$ as function of radius and 
time for the two PN simulations.}
\label{PNmodels}
\end{figure*}

To study the effects of H-deficient [WC]-type winds on the formation of PNe,
we use the approach from \citet{KahnBreitschwerdt} and
\citet{DwarkadasBalick}.  The evolution of the stellar wind in the post-AGB
phase is not well known, not for normal, nor for [WC] central stars. Hence a
simple power law behaviour is postulated. Following \citet{DwarkadasBalick}
the fast wind evolves according to
\begin{equation}
  v_{\rm fw}=v_{\rm fw}(0)\left(1+{t\over \tau}\right)^\alpha
  \label{timedep}
\end{equation}
The initial velocity $v_{\rm fw}(0)$ is taken to be 25~km~s$^{-1}$,
$\alpha=1.5$, and $\tau$ is chosen such that at $t=3000$~years, $v_{\rm fw}=
2000$~km~s$^{-1}$.  The mass loss changes so that the product $\dot{M}_{\rm
fw}v_{\rm fw}$ remains constant. This is expected for a radiation-driven wind
(of constant efficiency).

The environment is taken be a slow AGB wind characterized by a constant mass
loss rate of $2 \times 10^{-5}$~M$_\odot$~yr$^{-1}$ and a velocity of
10~km~s$^{-1}$. Its abundances are solar.

We calculated the evolution of this system for two cases, one for a fast wind
with normal, solar abundances, and one for a fast wind with [WC]
abundances. These abundances are as listed in Table~1. The wind speed
evolution is the same in both models, but the mass loss rate is taken to be 10
times higher in the [WC] case. The initial values (at a velocity of
25~km~s$^{-1}$) are $10^{-5}$~M$_\odot$~yr$^{-1}$ for the [WC] star, and
$10^{-6}$~M$_\odot$~yr$^{-1}$ for the H-rich star. As pointed out above, the
mass loss rate drops as the velocity increases. At the end of the calculation
when the wind has a velocity of 2000~km~s$^{-1}$, the mass loss rates are
$10^{-7}$ and $10^{-8}$~M$_\odot$~yr$^{-1}$, respectively.

Figure~\ref{PNmodels} shows the evolution of the logarithm of the density as
function of radius and time. Light shades correspond to higher densities. In
this representation the evolution of the shell is easily traced. In both cases
the bubble is initially momentum-driven and the inner shock and contact
discontinuity lie close to each other, at the inner edge of the swept up
shell. In the case of a fast wind with solar abundances, the inner shock
detaches itself from the contact discontinuity at a time of 450~years, when
the stellar wind has a velocity of 150~km~s$^{-1}$, and the shell has reached
a radius of $7.5\times 10^{-2}$~pc.  This is in line with what was found by
previous authors \citep{KahnBreitschwerdt, DwarkadasBalick}. In the case of
the [WC] star, the transition happens at a much later time, 1100 years after
the AGB, when the wind speed has reached 480~km~s$^{-1}$, at a radius of
$2.2\times 10^{-1}$~pc. Since the cooling time is proportional to the third
power of the shock velocity, and inversely proportional to the density, this
implies that the cooling is effectively 30 times higher (the effect of the
density is largely cancelled due to the fact that the transition happens at a
larger radius).

With the wind velocities mentioned above, we are reaching the regime where
other elements such as Fe, can contribute substantially to the cooling. This
could mean that the actual transition velocity for the solar abundances case
is somewhat higher. However, our value is in line with what was found before,
even with cooling curves including these heavier elements. Apparently, our
omission is not so serious for this case.  For the [WC] case, where the
abundances of the heavier elements are mostly `normal', their contribution
to the cooling is negligible compared to highly overabundant carbon.

The expansion velocity of the PN shell is found to be 50\% higher in the case
of the [WC] star, e.g. 21~km~s$^{-1}$ versus 17~km~s$^{-1}$ at
$t=2000$~years.  This is mostly due to the higher mass loss rate in
the [WC] stellar wind.

We also ran simulations with a much steeper rise of the fast wind
velocity, using $\alpha=2.5$ in Eq.~\ref{timedep}. As was found in previous
work \citep{KahnBreitschwerdt,DwarkadasBalick}, this does not affect the
transition velocities. Obviously, the transition times do change. Since the
behaviour of the post-AGB wind is not well known, one cannot say much about
the actual transition time, only about the transition velocity.

This shows that the [WC] character does change the evolution of the SWB
during the early phases of PN formation: higher mass loss rates will lead to
higher expansion velocities and the high C abundance will increase the wind
velocity at which the SWB changes from momentum- to energy-driven. The
implications of this and the connection to observations are discussed in
Sect.~6.

\begin{table*}
{Table 2 -- Input parameters for the RNe models}\\
\label{RN-input}
\begin{tabular*}{11.5cm}{*{6}{l}}
\noalign{\smallskip}
\hline\hline
\noalign{\smallskip}
                     & Solar & WNl   &WNe     & WCe    & WCl\\
\noalign{\smallskip}
\hline
\noalign{\smallskip}
$\dot{M}_{\rm  sw}$ (\myr)  & $8\times 10^{-5}$ & $8\times 10^{-5}$ 
                            & $8\times 10^{-5}$ & $8\times 10^{-5}$ 
                            & $8\times 10^{-5}$\\
$v_{\rm sw}$ (\kms)         & 30    & 30    & 30     & 30     &  30\\
$\cal{M}_{\rm sw}$ (mach number)         & 1.1  & 1.1  & 1.1  & 1.1 & 1.1\\
$\dot{M}_{\rm fw}$ (\myr)   & $1.5\times 10^{-5}$ & $3\times 10^{-5}$ 
                            & $3\times 10^{-5}$   & $1.5\times 10^{-5}$  
                            & $1\times 10^{-5}$\\
$v_{\rm fw}$ (\kms)         & 2300 & 840   & 2060   & 2300   & 1400\\
$\cal{M}_{\rm fw}$ (mach number)         & 20  & 20  & 20  & 20 & 20\\
$r_0$ (cm)                  & $5\times 10^{15}$ & $5\times 10^{15}$  
                            & $5\times 10^{15}$ & $5\times 10^{15}$ 
                            & $5\times 10^{15}$\\
grid size                   & 3200 & 3200 & 3200 & 3200 & 3200\\
            \noalign{\smallskip}
            \hline
\end{tabular*}
\end{table*}

\begin{figure*}
\centerline{\includegraphics[width=8cm]{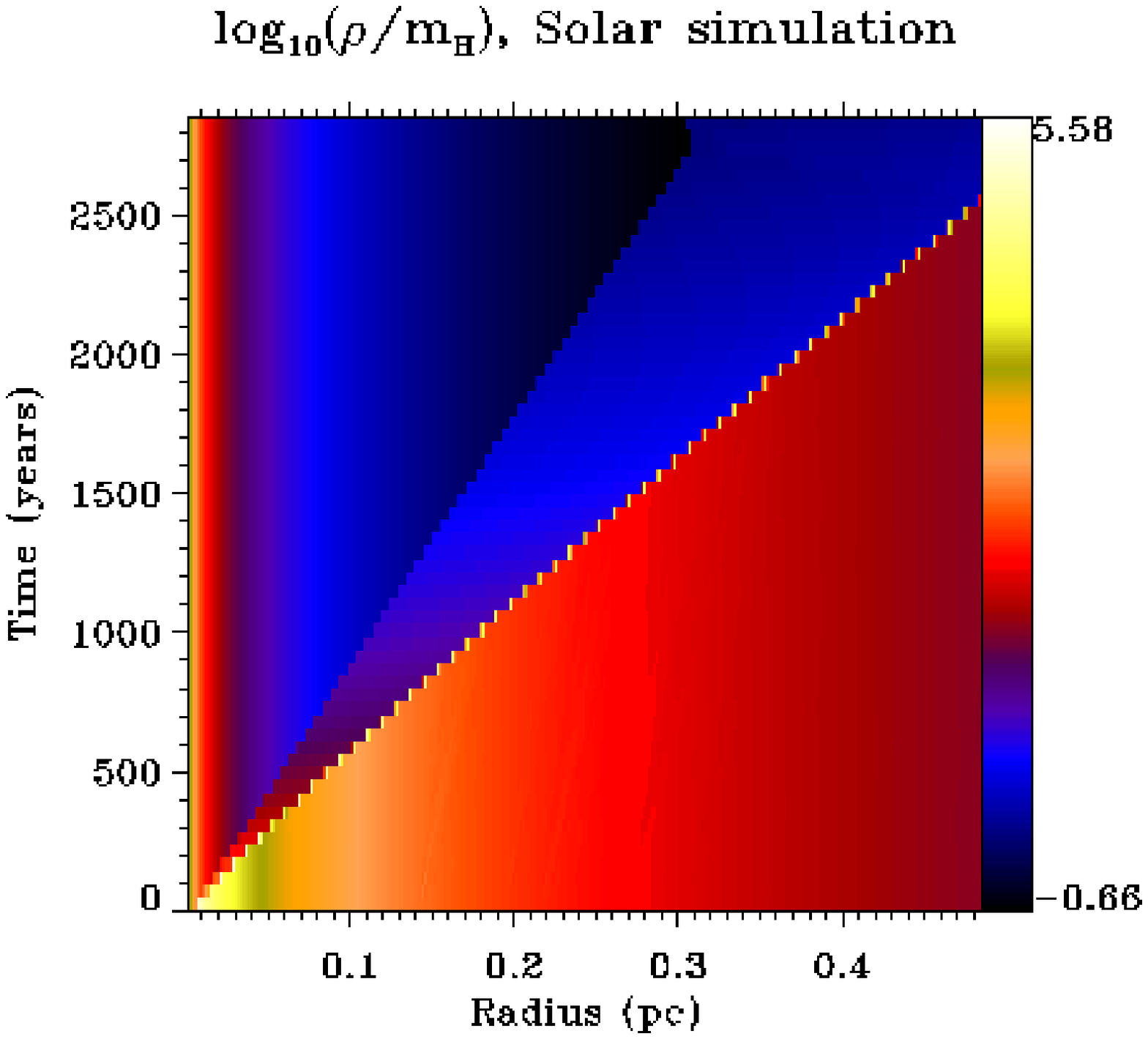}\includegraphics[width=8cm]{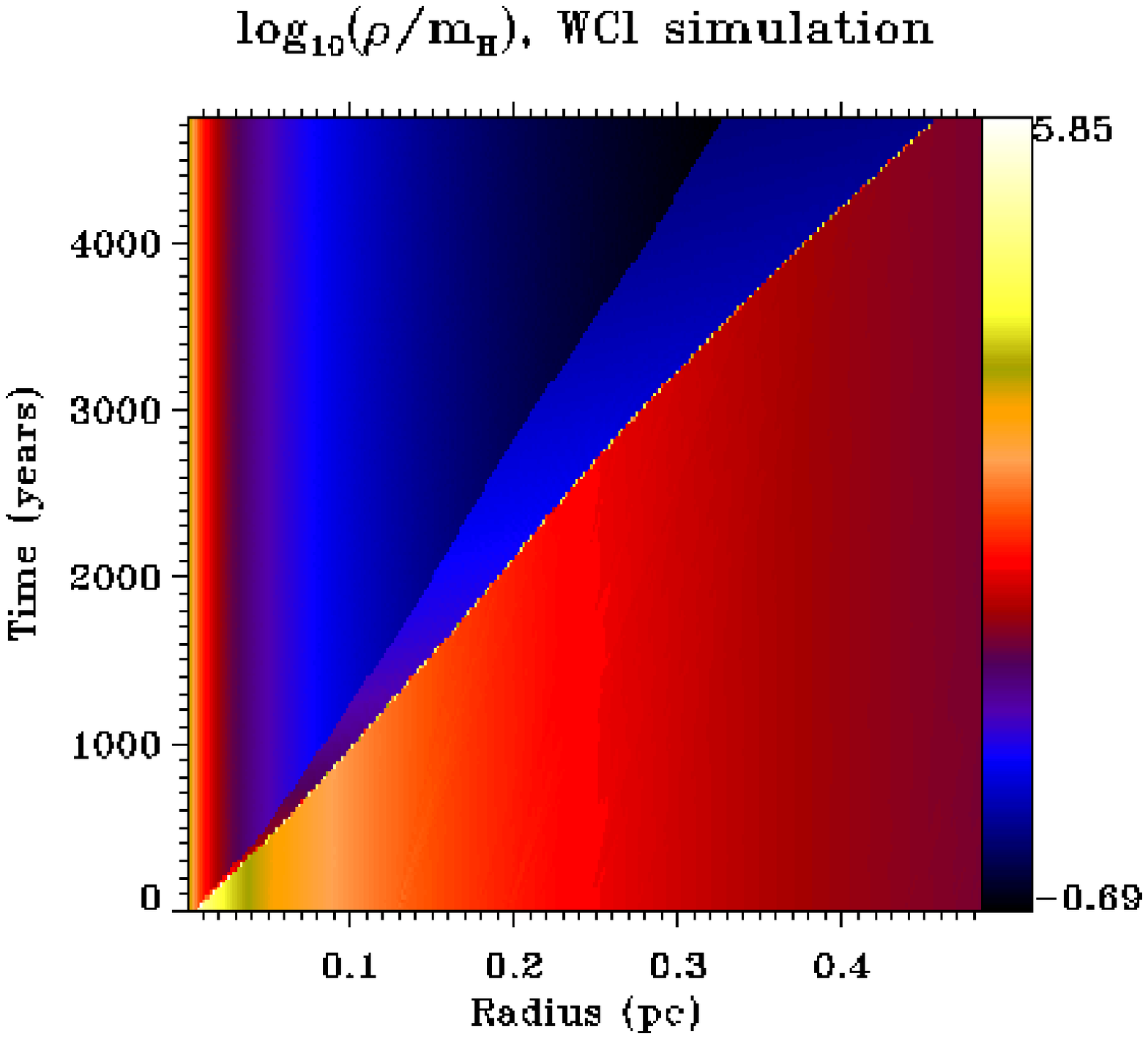}}
\centerline{\includegraphics[width=8cm]{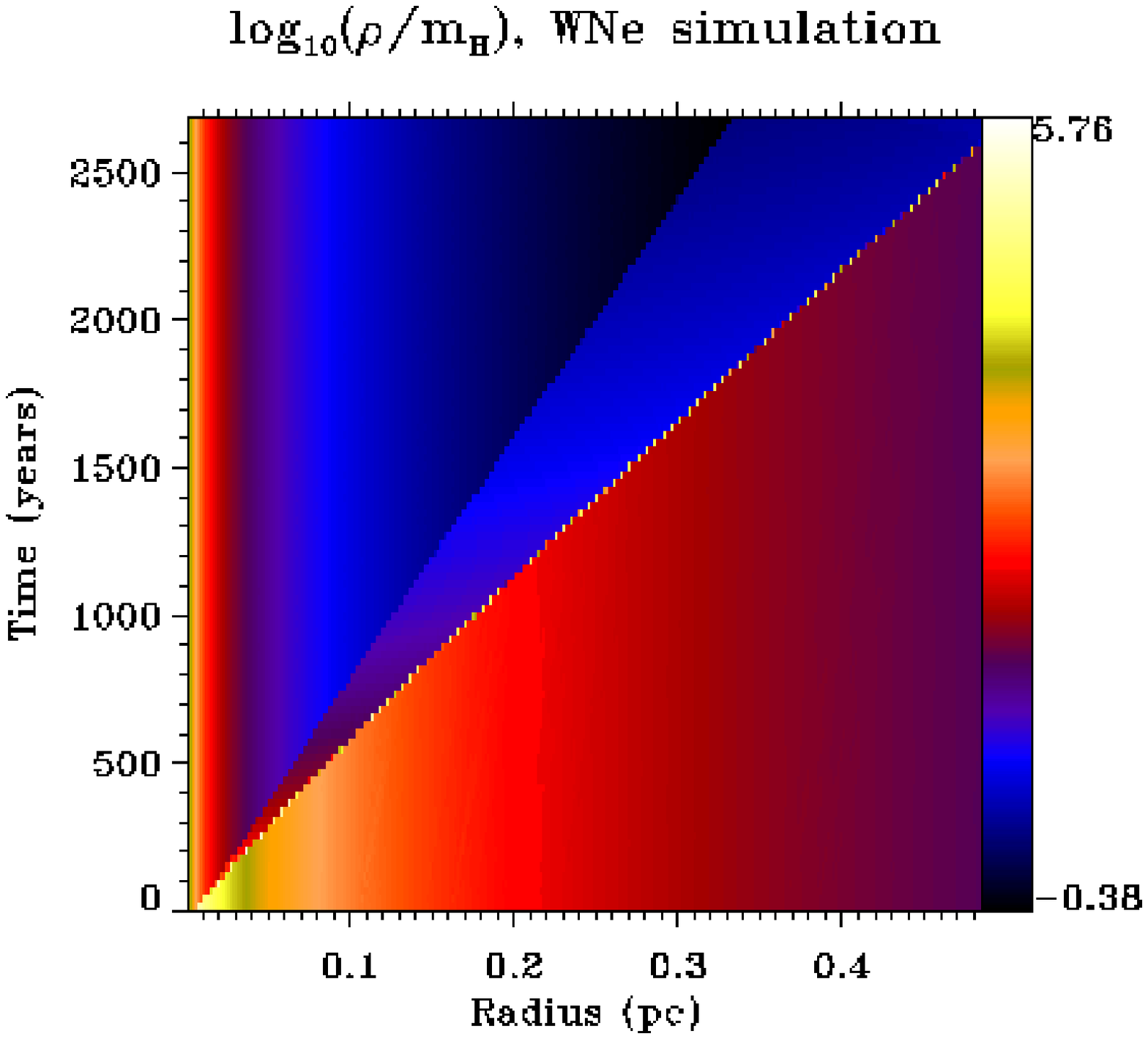}\includegraphics[width=8cm]{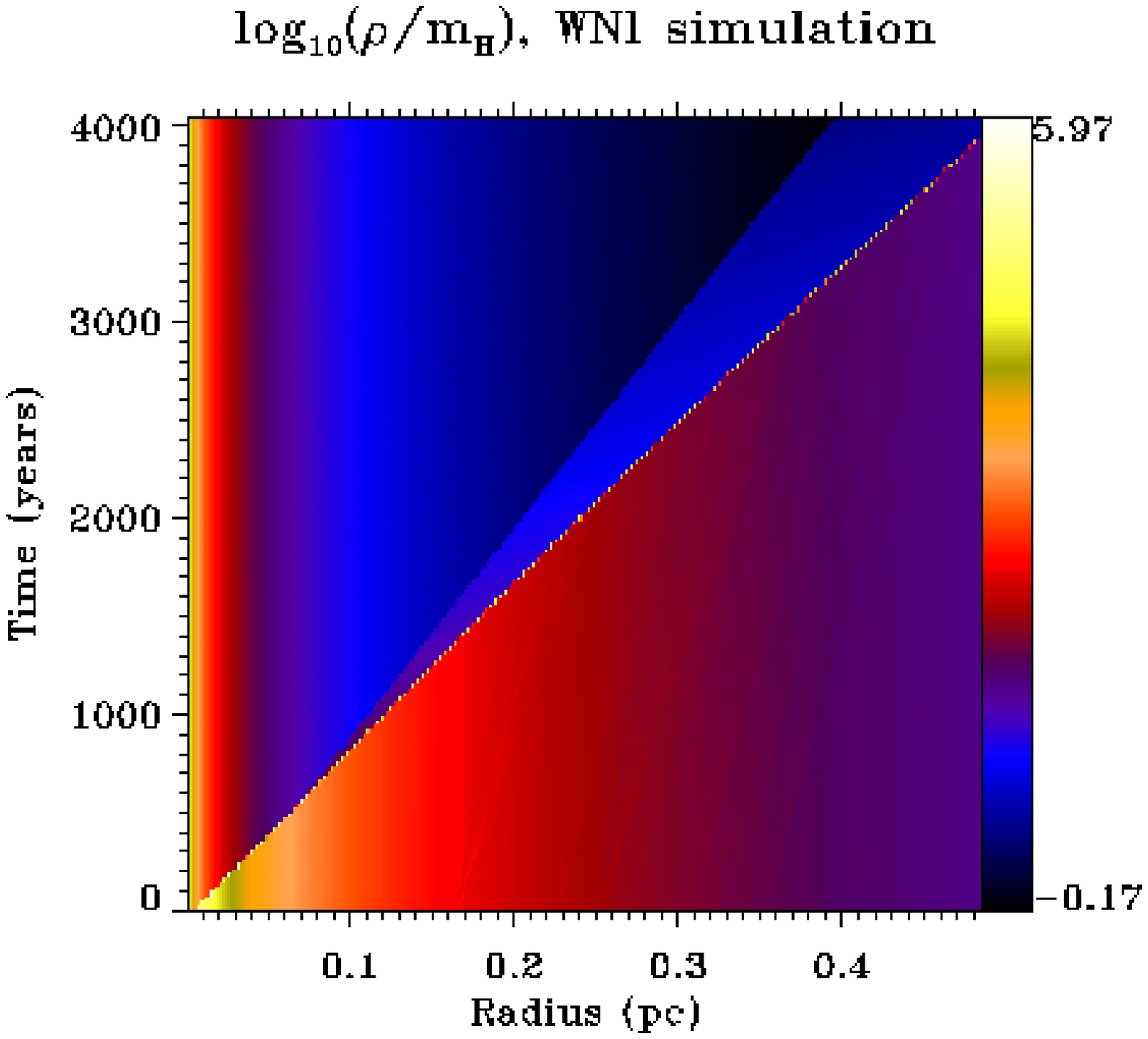}}
\caption{Colour plot of $\log_{10}(\rho /m_{\rm H})$ as function of radius and 
time for four of the five RN simulations.}
\label{RNmodels}
\end{figure*}

\section{Models for RNe}

For the RNe it is less easy to define one general model. First of all, the
different types of WR stars have different abundances (see Table~1), where one
distinguishes between nitrogen-rich WN and carbon-rich WC. Largely depending
on the effective temperature, the subclass `l' for late, i.e.~cooler, and `e'
for early, i.e.~hotter can be added.

The results from the previous section show that for WC abundances, a large
effect can be expected. However, for the WN stars, the metals are an order of
magnitude less abundant, helium being by far the most abundant element. Some
stars are thought to evolve from WN into WC, so the abundances in those WR
winds will change with time. The other complication is the mass loss history
of these stars, as pointed out in the introduction.

In order to keep the numerical experiment simple, we run simulations of a WR
wind running into a Red Super Giant (RSG) wind, keeping the abundances in the
WR wind constant. This problem then resembles that of the PNe, except that the
mass loss rates are higher, and the wind speed is constant in time. This way
we can study the effects of the H-deficient cooling on the bubble evolution.

We ran five simulations which only differed in the abundance of the stellar
wind, one with solar abundances, and one for each type of WR star: WNl, WNe,
WCl, WCe. All parameters are as listed in Table~2.

Figure~\ref{RNmodels} shows the evolution of the logarithm of the number
density in the same way as in Fig.~\ref{PNmodels}. One notices how the lowest
velocity wind (WNl) model produces a momentum-driven RN until $t=550$~years,
at a radius $7.2\times 10^{-2}$~pc, after which it switches to energy-driven.
The `solar', WNe, WCe, and WCl models are all energy-driven from the start,
as is the WCe model (not shown).

Due to the low densities, collisional ionization at the inner shock takes 
some time, and there is transition zone where the ionic concentrations are
out of equilibrium. This zone is quite narrow ($2\times 10^{16}$~cm), and
its existence does not have a major effect on the bubble dynamics or expected
X-ray spectrum.

\subsection{Momentum and energy ratios}
\cite{LamersCassinelli} briefly discuss the dynamics of SWBs in
Chap.~12. There they point out that one can determine whether a SWB is
momentum-driven or energy-driven by estimating the momentum and kinetic
energy of the swept-up shell, and comparing this to the momentum and kinetic
energy supplied by the stellar wind during the lifetime of the SWB. The ratio
of shell momentum to wind momentum is called $\pi$, the ratio of the kinetic
energies is $\epsilon$.
\begin{eqnarray}
  \pi  &=& {M_{\rm shell} v_{\rm shell} \over \dot{M}_{\rm fw}v_{\rm fw} t}\\
  \epsilon &=& {M_{\rm shell} v_{\rm shell}^2 \over \dot{M}_{\rm fw}v_{\rm
  fw}^2 t}\,.
  \label{etapi}
\end{eqnarray}

\citet{GarciaSeguraMacLow95a} derived an analytical description for SWBs
around WR stars, including expressions for the ratios $\pi$ and
$\epsilon$. However, they assumed that $v_{\rm sw}$ can be taken to be
zero. While it is true that for WR stars $v_{\rm sw} \ll v_{\rm fw}$, often
$\dot{M}_{\rm fw} v_{\rm fw} \sim \dot{M}_{\rm sw} v_{\rm sw}$, meaning that
one should take the original momentum of the slow wind into account. Also,
$v_{\rm sw}$ can be a sizeable fraction of $v_{\rm shell}$, which means that
the original velocity of the slow wind can be a non-negligible fraction of the
shell velocity.

Using a thin-shell approximation, the expansion velocity of the shell can be 
found by considering momentum- and energy conservation. For an energy-driven
bubble, momentum conservation means that
\begin{equation}
  {d \over dt}\left[M_{\rm shell}(\dot{r}-v_{\rm sw})\right]=4\pi r^2 P\,,
  \label{momedr}
\end{equation}
where $P$ is the pressure of the hot bubble. This pressure is related to 
input wind energy through energy conservation
\begin{equation}
  {d \over dt}\left({4\pi r^3 \over 3}{P \over (\gamma -1)}\right) =
   {1\over 2} \dot{M}_{\rm fw} v_{\rm fw}^2 - 4\pi r^2 P {dr\over dt}\,,
\end{equation}
where the radius of the inner shock is assumed to be much smaller than 
the radius of the shell.
The mass of the swept up shell ($M_{\rm shell}$) at a time $t$, when the shell
has reached a radius $r=v_{\rm exp}t$, is given by
\begin{eqnarray}
  M_{\rm shell}&=&\int_{0}^{r-v_{\rm sw}t} 4\pi r^2 \rho_0 {r_0^2\over r^2} dr
  \nonumber\\
            &=&4\pi r_0^2\rho_0(r-v_{\rm sw}t)={\dot{M}_{\rm sw}\over 
             v_{\rm sw}}(r-v_{\rm sw}t)\,,
  \label{mshell}
\end{eqnarray}
where $\dot{M}_{\rm sw}=4\pi r_0^2 \rho_0 v_{\rm sw}$, with $r_0$ the 
radius at which the slow wind density is $\rho_0$.

Equations \ref{momedr}--\ref{mshell} can be combined into one equation of
motion for the shell
\begin{equation}
  {\dot{M}_{\rm sw}\over 4v_{\rm sw}}{d \over dt}\left[r^3{d^2 \over dt^2}
    (r-v_{\rm sw}t)^2\right]={1\over 2} \dot{M}_{\rm fw} v_{\rm fw}^2
\end{equation}
The solution of this equation is a shell expanding with constant velocity,
\begin{eqnarray}
  \lefteqn{v_{\rm exp,e} = v_{\rm sw} + }\nonumber\\
  &&v_{\rm *,e}
 \Bigg[
  \left( {1 \over 2}
     \left( 1+
        \sqrt{1-4\left( {v_{\rm sw} \over 3v_{\rm *,e}} \right)^3}
     \right) - 
     \left({v_{\rm sw} \over 3v_{\rm *,e}}\right)^3
   \right)^{1/3} + \nonumber\\
  &&\left({1 \over 2}
    \left(1-\sqrt{1-4
       \left({v_{\rm sw} \over 3v_{\rm *,e}}\right)^3}
    \right) - 
       \left({v_{\rm sw} \over 3v_{\rm *,e}}\right)^3
   \right)^{1/3}
 \Bigg]\,,
     \label{vexpe}
\end{eqnarray}
where 
\begin{equation}
  v_{\rm *,e} = \left({1\over 3} {\dot{M}_{\rm fw} \over
     	\dot{M}_{\rm sw}} v_{\rm fw}^2 v_{\rm sw} \right)^{1/3}\,,
\end{equation}
the solution for inifinitely small $v_{\rm sw}$, which can be found in
for example \citet{GarciaSeguraMacLow95a} (Eq.~30).
Neglecting terms of order $(v_{\rm sw}/3v_{\rm *,e})^3$, Eq.~\ref{vexpe} 
can be approximated as 
\begin{equation}
  v_{\rm exp,e} \simeq v_{\rm sw} + v_{\rm *,e}\,.
  \label{approxvexpe}
\end{equation}

For the momentum conserving case it is easier to consider the balance
between input momentum and shell momentum. The shell now consists of
compressed material from both the slow and fast winds. The mass of swept up
slow wind material which we will call $M_{\rm sh,sw}$ is still given by
Eq.~\ref{mshell}, the mass of the fast wind material compressed into a shell
at a position $r$ at a time $t$ is given by
\begin{equation}
  M_{\rm sh,fw} = {\dot{M}_{\rm fw}\over v_{\rm fw}}(v_{\rm fw}t-r)\,,
\label{mshfw}
\end{equation}
i.e.~the amount of mass supplied by the fast wind in a time $t$, minus the
material which has not joined the shell yet.

Conservation of momentum says that the current momentum of the shell
should equal the momentum added by the fast and slow winds:
\begin{equation}
  (M_{\rm sh,fw}+M_{\rm sh,fw})v_{\rm sh} =
  M_{\rm sh,fw}v_{\rm fw}+M_{\rm sh,fw}v_{\rm sw}\,.
\end{equation}
Substituting Eqs.~\ref{mshell} and \ref{mshfw} into this expression and some
algebra, shows that the solution is a shell with constant expansion velocity
\begin{eqnarray}
  v_{\rm exp,m}  = \frac{v_{\rm sw}+v_{\rm *,m}}{1+v_{\rm *,m}/v_{\rm
      fw}}\,,
  \label{vexpm}
\end{eqnarray}
where
\begin{equation}
  v_{\rm *,m}\equiv\sqrt{{\dot{M}_{\rm fw} \over
    \dot{M}_{\rm sw}} v_{\rm fw} v_{\rm sw}}\,,
\end{equation}
the solution for ${\dot{M}_{\rm fw}/\dot{M}_{\rm sw}} \to 0$
(disregarding the mass contribution of the fast wind), and infinitely
small $v_{\rm sw}$ (disregarding the slow wind's momentum input). This
was the expression effectively used by \citet{GarciaSeguraMacLow95a}.

Only dropping the terms linear or higher in
${\dot{M}_{\rm fw}/\dot{M}_{\rm sw}}$ and $v_{\rm sw}/ v_{\rm fw}$, one
obtains the approximate solution
\begin{equation}
  v_{\rm exp,m}\simeq v_{\rm sw}+v_{\rm *,m}\,,
\label{approxvexpm}
\end{equation}
\citep[see e.g.][Eq.~16]{KahnBreitschwerdt}

Using these new expressions for the expansion velocities, Eqs.~\ref{vexpe}
and \ref{vexpm}, and assuming a radiative outer shock so that $v_{\rm
shell}=v_{\rm exp}$, one can easily derive general expressions for the ratios
$\pi$ and $\epsilon$. Since the expression for the expansion velocity in
the energy-driven case is quite complicated, we will use the approximate
expression Eq.~\ref{approxvexpe} to derive a first order approximation for
$\pi$ and $\epsilon$ in case of a non-zero slow wind velocity. This solution
is valid as long as $\dot{M}_{\rm fw}v_{\rm fw}^2\ll \dot{M}_{\rm sw}v_{\rm
sw}^2$. We thus obtain
\begin{eqnarray}
  \pi_{\rm e} &\simeq& {1 \over 3}{v_{\rm fw} \over v_{\rm *,e}}+
                  {1 \over 3}{v_{\rm fw}v_{\rm sw} \over (v_{\rm *,e})^2}
	\label{pie}
	\\
  \epsilon_{\rm e} &\simeq& {1 \over 3} + 
        {1 \over 3}{v_{\rm sw}\over v_{\rm *,e}}
	\left(2+{v_{\rm sw}\over v_{\rm *,e}}\right)\,.
	\label{epse}
\end{eqnarray}
For the momentum-driven case, the exact ratios are
\begin{eqnarray}
  \pi_{\rm m} &=& \frac{v_{\rm fw}}{v_{\rm fw}+v_{\rm *,m}}+
    v_{\rm sw}\frac{v_{\rm fw}-
    v_{\rm *,m}-v_{\rm sw}}{v_{\rm *,m}v_{\rm fw}+v_{\rm *,m}^2}
  \label{pim}
  \\ \epsilon_{\rm m} &&= \frac{v_{\rm fw}v_{\rm *,m}}{(v_{\rm fw}+v_{\rm
  *,m})^2} + \nonumber\\
  &&v_{\rm sw}\frac{v_{\rm *,m}(2v_{\rm fw}-v_{\rm *,m})+v_{\rm
  sw}(v_{\rm fw}-2v_{\rm *,m}-v_{\rm sw})}{v_{\rm *,m}(v_{\rm fw}+v_{\rm
  *,m})^2}
	  \label{epsm}
\end{eqnarray}
Taking $v_{\rm sw}=0$, and dropping terms of order $v_{\rm *}/v_{\rm fw}$,
one can recover the expressions from \citet{GarciaSeguraMacLow95a}, their
Eqs.~83 and 87.

According to \citet{LamersCassinelli}, RNe are mostly momentum-driven. This is
mostly based on the work of \citet{TreffersChu82} and \citet{Chu82}, who
derived momenta and kinetic energies for a set of RNe, and found small values
for $\pi$ and $\epsilon$, typically $\pi\sim 0.5$ and $\epsilon \sim 0.01$. 
More recent work, for example of \citet{Cappaetal96}, gives similar results
for other nebulae.

\citet{vanBuren} recalculated $\pi$ and $\epsilon$ for the five RNe from
\citet{Chu82}. He used the same data, but applied a correction for an unseen
neutral component. He then finds larger values for both $\pi$ and $\epsilon$:
$\pi$ ranges from 1.2 to 6.8; $\epsilon$ ranges from 0.013 to 0.15. These
values would make the SWBs energy-conserving rather than momentum-conserving.

Observationally, one can think of a number of other complications when
determining $\pi$ and $\epsilon$, such as determination of the mass and
velocity of a clumpy flow, non-sphericity effects, unknown mass loss history,
estimating the age of the RN, etc. As shown above, if the RN is expanding
into material from a previous slower wind, one would also need an estimate of
its velocity $v_{\rm sw}$.

This raises the question whether observationally derived values for
$\pi$ and $\epsilon$ can be used to derive the character of the RNe.

\begin{table}
{Table 3 -- Momentum and kinetic energy ratios for RNe models}\\
\begin{tabular*}{9cm}{*{6}{l}}

\noalign{\smallskip}
\hline\hline
\noalign{\smallskip}
                    & Solar & WNl   & WNe  & WCe   & WCl  \\
\noalign{\smallskip}
\hline
\noalign{\smallskip}
$\pi$               & 2.4   & 1.4   & 1.5  & 1.02  & 0.9  \\
$\pi_{\rm e}$       & 4.1   & 2.5   & 3.1  & 4.1   & 4.2  \\
$\pi_{\rm m}$       & 1.2   & 1.1   & 1.1  & 1.2   & 1.3  \\
\noalign{\smallskip}
$\epsilon$          & 0.21  & 0.20  & 0.16 & 0.056 & 0.06 \\
$\epsilon_{\rm e}$  & 0.43  & 0.49  & 0.42 & 0.43  & 0.50 \\
$\epsilon_{\rm m}$  & 0.07  & 0.15  & 0.09 & 0.07  & 0.09 \\
\noalign{\smallskip}
\hline
\end{tabular*}
\end{table}

In order to better evaluate the observationally derived values, we extract
values of $\pi$ and $\epsilon$ from our simulations.  In order to check that
our procedure works, we determined $\pi$ and $\epsilon$ for two simulations in
which the radiative source terms were switched off, but which used adiabatic
indices $\gamma=5/3$ (energy-conserving) and $\gamma=1.001$
(momentum-conserving), respectively. The results from these simulations
reproduce the analytical values from Eqs.~\ref{pie}--\ref{epsm} within 5\%.

In Table~3 we list the values found from the RN simulations, and the one
expected from Eqs.~\ref{pie}--\ref{epsm}. What one sees is that even though
the SWBs are all energy-driven, in the sense that they contain hot shocked
fast wind material, the values are lower than expected. This means that the
partially radiative character of the SWB reduces the values for $\pi$ and
$\epsilon$, and approach the values for momentum-driven bubbles. For the two
WC simulations, the value actually drops below the ones expected for the
momentum-driven case. This is because the swept-up material cools below its
pre-shock temperature. The cooling is more efficient here, due to mixing in of
wind material into the shell.

These results shows that the ratios $\pi$ and $\epsilon$ are of very limited
use when establishing the character of SWBs: an energy-driven bubble may
disguise itself as a momentum-driven bubble.

\section{Discussion}

\subsection{PNe}
The effects of the [WC] wind in PNe may seem limited, only postponing
the transition from momentum- to energy-driven bubbles to a later
time. However, this later transition may have some serious consequences. First
of all, during the momentum-driven phase the (proto-)PN suffers from the
non-linear thin shell instability, \citep[see e.g.][] {DwarkadasBalick}. A
later transition gives this instability a longer time to work on the structure
of the nebula. The result could then be a more disturbed or fragmented
nebula. There are some observational indications for more substructure in
[WR]-PNe \citep{Gorny2001}. More quantitatively, analysis of the emission
lines from a number of [WR]-PNe shows that these lines are quite wide,
sometimes wide enough to hide any splitting of the lines \citep{Acker2002}. To
reproduce these line shapes, the authors invoke a turbulent velocity
component, something which is not needed for normal PNe. This could be an
effect of the later transition. 

Because of the width of the lines, finding the expansion velocity of the
nebulae becomes somewhat difficult. Earlier results indicated higher than
average expansion velocity for [WR]-PNe \citep{Gorny2001}, but
\citet{Acker2002} dispute this. Our models do predict a higher average
expansion velocity, mostly due to the higher mass loss rates. Of course there
will be large variations, depending on the precise mass loss history of the
star, the age of the nebula, and the lines used to measure the expansion
velocity.

If the transition from momentum- to energy-driven takes longer for [WR]-PNe,
there is another interesting effect. During the momentum-driven phase,
asphericities in the fast wind are easily imprinted on the swept-up shell. As
the hot bubble develops, it forms more and more of a buffer between angular
variations in the fast wind and the swept-up shell, and the aspherical fast
wind has less and less of an effect on the shape of the nebula. If PNe acquire
most of their asphericity due to an aspherical fast wind, one would therefore
expect [WR]-PNe, which develop a hot bubble later, to be more aspherical.
Since there seems to be no systematic morphological differences between
[WR]-PNe and normal PNe \citep{Gorny2001}, the conclusion would be that
aspherical post-AGB winds do not play a major role in determining the
asphericity of PNe. This rather qualitative statement will be studied in some
more detail in a subsequent paper.

The fact that the [WR]-PNe form a subgroup with clearly different mass loss
properties may allow further tests of models for the formation of PNe in
general. For instance, the efficiency of the magnetic shaping mechanism of
\citet{ChevalierLuo} is given by
\begin{equation}
  \sigma = {B_{\rm star}^2 R_{\rm star}^2 \over \dot{M} v_{\rm fw}}
    \left({v_{\rm rot} \over v_{\rm fw}}\right)^2\,,
\end{equation}
where $B_{\rm star}$ is the magnetic field strength at the surface of the
star, $R_{\rm star}$ is the stellar radius, $v_{\rm rot}$ is the stellar
rotation velocity. If typical field strengths and rotation speeds for [WR]
central stars are the same as for normal central stars, the higher mass loss
rates for [WR] stars implies lower values of $\sigma$, i.e.~lesser efficiency
of the magnetic shaping. This would then give on average rounder [WR]-PNe. The
magnitude of the effect is about a factor 10. The uncertainties in the other
key parameters ($B_{\rm star}$, $v_{\rm rot}$), may make it hard to find
this effect in the observations. However, in all models the formation of a PN
is basically driven by the post-AGB wind, and that there are clear differences
between the winds of [WC] and normal central stars, a closer comparison of the
shapes of [WC]-PNe to normal PNe is warranted.

\subsection{RNe}
The numerical models presented in the previous section show that it is
unlikely that RNe around WR stars are momentum-driven. The amount of energy
produced in shocking the fast wind cannot efficiently be radiated away, not
even by the extremely metal-rich gas of the WR wind. Unless there are other
physical mechanisms at work, the RNe cannot be momentum-driven. 

A number of physical mechanisms may alter the results of our simplified
models. Photo-ionization will alter the ionization state of parts of the
RNe. However, the spectrum of WR stars is not hard enough to produce the ions
which are collisionally generated in the hot bubble, and hence it will not
change our conclusion about the energy-driven character of RNe. Clumpiness and
time variability of the stellar wind will produce areas of different
temperatures when shocked. However, the lower density portions of the wind
will cool slowly, and are likely to take up the largest volume, so the bubble
will still be filled with a substantial amount of hot gas, making it
energy-driven. Clumpiness in the environment is likely to be more effective in
modifying the character of the RNe. If these clumps are not effectively
accelerated by the outer shock, they may end up inside the SWB, and start to
evaporate, raising the density, and lowering the temperature of the hot
shocked wind material, a process referred to as `mass loading'. This could
conceivably lead to more efficient cooling and a collapse of the hot bubble
against the inner edge of the shell. Thermal conduction, if not inhibited by
the presence of a magnetic field, can produce a similar result. However,
studies of mass loading \citep{Pittard} show no indication for such a 
collapse.

The observed X-ray spectra for RNe show the presence of hot gas
\citep{NGC6888X,S103X}, although not as hot as predicted by the models.
This is a strong observational indication that RNe are filled with hot gas,
and are therefore energy-driven.

The remaining question is how to interpret the observationally derived values
for $\pi$ and $\epsilon$. Our simulations show that these are of limited
value. Even in the idealized case studied here, $\pi$ and $\epsilon$ do not
follow the simple division between momentum- and energy-driven. The
observational problems are also many, as was already outlined by
\citet{vanBuren}. In addition the complex and fast evolution of these stars
and their winds, invalidates the assumption that one can simply integrate over
time. The momentum and kinetic energy of the shells are an integration over
the momenta and kinetic energy of the swept-up material, which could come from
the interstellar medium and various previous mass loss phases. Dividing this
by a value derived from assuming the current wind to have been constant in
time, is unlikely to give sensible values.

\section{Conclusions}
We simulated the effects H-deficient winds have on their SWBs, studying
simplified cases for PNe and RNe. We find that the extreme abundances in the
winds of [WC] stars can keep their PNe momentum-driven for a longer time. We
speculate that this leads to more turbulent nebulae and would also produce more
aspherical PNe if their shape was mostly due to an aspherical post-AGB wind.

For the RNe around massive WR stars, we showed that they cannot be
momentum-driven, despite earlier claims to the opposite. We pointed out some
of the difficulties related to the deriving the momentum- versus
energy-driven character of the RNe using the comparisons of RN momentum and
kinetic energy to the assumed total input of momentum and kinetic energy by
the stellar wind.

The models in this article illustrate the physical effects of abundances on
the structure of SWBs. To be realistic models for PNe and RNe, they should be
improved in several ways. As shown for instance in \citet{RHPNIII},
ionization fronts can play an important role in the early evolution of
PNe. For the RNe adding photo-ionization to the models could help in
estimating the amount of neutral material, which is one of the unknowns in
the $\pi$-$\epsilon$ method. Letting their winds evolve according to a more
realistic recipe such as in \citet{RHPNIII} for PNe, and in 
\citet{GuileNorbert2} for RNe, would be another step in the direction of
more realistic models. However, given that the results in this paper show 
that for RNe the wind abundances play only a minor role, the models of
\citet{GuileNorbert1} and \citet{GuileNorbert2} remain largely valid, and
only for the case of the [WR]-PNe it makes sense to pursue more realistic
models.

\begin{acknowledgements}
This work was sponsored by the Stichting Nationale Computerfaciliteiten
(National Computing Foundation, NCF) for the use of supercomputer facilities,
with financial support from the Nederlandse Organisatie voor Wetenschappelijk
Onderzoek (Netherlands Organization for Scientific Research, NWO).
The research of GM has been made possible by a fellowship of the Royal
Netherlands Academy of Arts and Sciences.
  PL is a Reserach Fellow at the Royal Swedish Academy supported by a grant 
from the Wallenberg Foundation.
GM and PL acknowledge support from the Swedish Research Council for funding
working visits between Stockholm and Leiden.
We thank the referee (R. Chevalier) for useful comments and pointing out an
error in the original version of this paper.
\end{acknowledgements}

\bibliographystyle{aa}
\bibliography{bubbles.bib}

\begin{thebibliography}{41}
\expandafter\ifx\csname natexlab\endcsname\relax\def\natexlab#1{#1}\fi

\bibitem[{{Abbott} \& {Conti}(1987)}]{AbbottConti87}
{Abbott}, D.~C. \& {Conti}, P.~S. 1987, \araa, 25, 113

\bibitem[{{Acker} {et~al.}(2002){Acker}, {Gesicki}, {Grosdidier}, \&
  {Durand}}]{Acker2002}
{Acker}, A., {Gesicki}, K., {Grosdidier}, Y., \& {Durand}, S. 2002, \aap, 384,
  620

\bibitem[{{Balick} {et~al.}(1993){Balick}, {Mellema}, \& {Frank}}]{Coolingfits}
{Balick}, B., {Mellema}, G., \& {Frank}, A. 1993, \aap, 275, 588

\bibitem[{{Cappa} {et~al.}(1996){Cappa}, {Niemela}, {Herbstmeier}, \&
  {Koribalski}}]{Cappaetal96}
{Cappa}, C.~E., {Niemela}, V.~S., {Herbstmeier}, U., \& {Koribalski}, B. 1996,
  \aap, 312, 283

\bibitem[{{Chevalier} \& {Luo}(1994)}]{ChevalierLuo}
{Chevalier}, R.~A. \& {Luo}, D. 1994, \apj, 421, 225

\bibitem[{{Chu}(1982)}]{Chu82}
{Chu}, Y.-H. 1982, \apj, 254, 578

\bibitem[{{Cox} \& {Daltabuit}(1971)}]{CoxDaltabuit71}
{Cox}, D.~P. \& {Daltabuit}, E. 1971, \apj, 167, 113

\bibitem[{{Cox} \& {Tucker}(1969)}]{CoxTucker}
{Cox}, D.~P. \& {Tucker}, W.~H. 1969, \apj, 157, 1157

\bibitem[{{Dalgarno} \& {McCray}(1972)}]{DalgarnoMcCray}
{Dalgarno}, A. \& {McCray}, R.~A. 1972, \araa, 10, 375

\bibitem[{{Dwarkadas} \& {Balick}(1998)}]{DwarkadasBalick}
{Dwarkadas}, V.~V. \& {Balick}, B. 1998, \apj, 497, 267

\bibitem[{{Dyson} \& {Williams}(1997)}]{DysonWilliams}
{Dyson}, J.~E. \& {Williams}, D.~A. 1997, {The physics of the interstellar
  medium, 2nd ed.} (Institute of Physics Publishing. ISBN 0750303069)

\bibitem[{{Frank} \& {Mellema}(1994)}]{RH-PN1}
{Frank}, A. \& {Mellema}, G. 1994, \aap, 289, 937

\bibitem[{{G{\' o}rny}(2001)}]{Gorny2001}
{G{\' o}rny}, S.~K. 2001, \apss, 275, 67

\bibitem[{{Gaetz} \& {Salpeter}(1983)}]{GaetzSalpeter}
{Gaetz}, T.~J. \& {Salpeter}, E.~E. 1983, \apjs, 52, 155

\bibitem[{{Garcia-Segura} {et~al.}(1996{\natexlab{a}}){Garcia-Segura},
  {Langer}, \& {Mac Low}}]{GuileNorbert2}
{Garcia-Segura}, G., {Langer}, N., \& {Mac Low}, M.-M. 1996{\natexlab{a}},
  \aap, 316, 133

\bibitem[{{Garcia-Segura} \& {Mac Low}(1995)}]{GarciaSeguraMacLow95a}
{Garcia-Segura}, G. \& {Mac Low}, M. 1995, \apj, 455, 145

\bibitem[{{Garcia-Segura} {et~al.}(1996{\natexlab{b}}){Garcia-Segura}, {Mac
  Low}, \& {Langer}}]{GuileNorbert1}
{Garcia-Segura}, G., {Mac Low}, M.-M., \& {Langer}, N. 1996{\natexlab{b}},
  \aap, 305, 229

\bibitem[{{Gould}(1980)}]{Gould80}
{Gould}, R.~J. 1980, \apj, 238, 1026

\bibitem[{{Gronenschild} \& {Mewe}(1978)}]{GronenschildMewe}
{Gronenschild}, E.~H.~B.~M. \& {Mewe}, R. 1978, \aaps, 32, 283

\bibitem[{{Hamann}(2002)}]{Hamann02}
{Hamann}, W.-R. 2002, in IAU Symp. 209: Planetary Nebulae, in press

\bibitem[{{Herwig}(2001)}]{Herwig2001}
{Herwig}, F. 2001, \apss, 275, 15

\bibitem[{{Kahn} \& {Breitschwerdt}(1990)}]{KahnBreitschwerdt}
{Kahn}, F.~D. \& {Breitschwerdt}, D. 1990, \mnras, 242, 505

\bibitem[{{Karzas} \& {Latter}(1961)}]{KarzasLatter}
{Karzas}, W.~J. \& {Latter}, R. 1961, \apjs, 6, 167

\bibitem[{{Koesterke}(2001)}]{Koesterke01}
{Koesterke}, L. 2001, \apss, 275, 41

\bibitem[{{Koo} \& {McKee}(1992{\natexlab{a}})}]{KooMcKee1}
{Koo}, B. \& {McKee}, C.~F. 1992{\natexlab{a}}, \apj, 388, 93

\bibitem[{{Koo} \& {McKee}(1992{\natexlab{b}})}]{KooMcKee2}
---. 1992{\natexlab{b}}, \apj, 388, 103

\bibitem[{{Lamers} \& {Cassinelli}(1999)}]{LamersCassinelli}
{Lamers}, H.~J.~G.~L.~M. \& {Cassinelli}, J.~P. 1999, {Introduction to stellar
  winds} (Cambridge University Press. ISBN 0521593980)

\bibitem[{{LeVeque}(2002)}]{Clawpack}
{LeVeque}, R.~J. 2002, {Finite Volume Methods for Hyperbolic Problems}
  (Cambridge University Press. ISBN 0521009243)

\bibitem[{{Mellema}(1994)}]{RHPNIII}
{Mellema}, G. 1994, \aap, 290, 915

\bibitem[{{Mellema} {et~al.}(1991){Mellema}, {Eulderink}, \&
  {Icke}}]{Mellemaetal91}
{Mellema}, G., {Eulderink}, F., \& {Icke}, V. 1991, \aap, 252, 718

\bibitem[{{Mellema} {et~al.}(1998){Mellema}, {Raga}, {Canto}, {Lundqvist},
  {Balick}, {Steffen}, \& {Noriega-Crespo}}]{Photoclump}
{Mellema}, G., {Raga}, A.~C., {Canto}, J., {et~al.} 1998, \aap, 331, 335

\bibitem[{{Pittard} {et~al.}(2001){Pittard}, {Hartquist}, \& {Dyson}}]{Pittard}
{Pittard}, J.~M., {Hartquist}, T.~W., \& {Dyson}, J.~E. 2001, \aap, 373, 1043

\bibitem[{{Plewa} \& {M{\" u}ller}(1999)}]{PlewaMuller}
{Plewa}, T. \& {M{\" u}ller}, E. 1999, \aap, 342, 179

\bibitem[{{Raga} {et~al.}(1997){Raga}, {Mellema}, \& {Lundqvist}}]{Ragaetal}
{Raga}, A.~C., {Mellema}, G., \& {Lundqvist}, P. 1997, \apjs, 109, 517

\bibitem[{{Raymond} {et~al.}(1976){Raymond}, {Cox}, \&
  {Smith}}]{RaymondCoxSmith}
{Raymond}, J.~C., {Cox}, D.~P., \& {Smith}, B.~W. 1976, \apj, 204, 290

\bibitem[{{Roe}(1981)}]{Roe81}
{Roe}, P. 1981, J.~Comp.~Phys, 43, 357

\bibitem[{{Sutherland} \& {Dopita}(1993)}]{DopitaSutherland}
{Sutherland}, R.~S. \& {Dopita}, M.~A. 1993, \apjs, 88, 253

\bibitem[{{Treffers} \& {Chu}(1982)}]{TreffersChu82}
{Treffers}, R.~R. \& {Chu}, Y.-H. 1982, \apj, 254, 569

\bibitem[{{Van Buren}(1986)}]{vanBuren}
{Van Buren}, D. 1986, \apj, 306, 538

\bibitem[{{Wrigge}(1999)}]{S103X}
{Wrigge}, M. 1999, \aap, 343, 599

\bibitem[{{Wrigge} {et~al.}(1998){Wrigge}, {Chu}, {Magnier}, \&
  {Kamata}}]{NGC6888X}
{Wrigge}, M., {Chu}, Y.-H., {Magnier}, E.~A., \& {Kamata}, Y. 1998, Berlin
  Springer Verlag Lecture Notes in Physics, v.506, 506, 425

\end{thebibliography}

\end{document}